\documentclass[pre,twocolumn,aps,eqsecnum]{revtex4}
\usepackage{epsfig}

\begin{document}

\title{Nonequilibrium Thermodynamics of the Kovacs Effect}

\author{Eran Bouchbinder}
\affiliation{Department of Chemical Physics, Weizmann Institute of Science, Rehovot 76100, Israel}

\author{J.S. Langer}
\affiliation{Department of Physics, University of California, Santa Barbara, CA  93106-9530}

\date{\today}

\begin{abstract}
We present a thermodynamic theory of the Kovacs effect based on the idea that the configurational degrees of freedom of a glass-forming material are driven out of equilibrium with the heat bath by irreversible thermal contraction and expansion.  We assume that the  slowly varying configurational subsystem, i.e. the part of the system that is described by inherent structures, is characterized by an effective temperature, and contains a volume-related internal variable.  We examine mechanisms by which irreversible dynamics of the fast, kinetic-vibrational degrees of freedom can cause the entropy and the effective temperature of the configurational subsystem to increase during sufficiently rapid changes in the bath temperature. We then use this theory to interpret the numerical simulations by Mossa and Sciortino (MS), who observe the Kovacs effect in more detail than is feasible in laboratory experiments. Our analysis highlights two mechanisms for the equilibration of internal variables. In one of these, an internal variable first relaxes toward a state of quasi-equilibrium determined by the effective temperature, and then approaches true thermodynamic equilibrium as the effective temperature slowly relaxes toward the bath temperature. In the other mechanism, an internal variable directly equilibrates with the bath temperature on intermediate timescales, without equilibrating with the effective temperature at any stage. Both mechanisms appear to be essential for understanding the MS results. 
\end{abstract}
\maketitle

\section{Introduction}
\label{intro}

The Kovacs effect reveals some of the most subtle and important nonequilibrium features of glassy dynamics. In particular, it provides detailed information about the ways in which glassy materials deform irreversibly and remember their histories of deformation \cite{KOVACS-63,KOVACS-79,MCKENNA-89,ANGELL-00}. Here, we develop a thermodynamic theory of the Kovacs effect, motivated in large part by the molecular-dynamics simulations of Mossa and Sciortino (MS)\cite{MS-04}.

In a Kovacs experiment, the volume of a glass-forming system is measured at fixed pressure as the temperature is varied.  A sample is first quenched from a high temperature $T_h$ to a temperature $T_{\ell}$ low enough -- i.e. near enough to the glass temperature $T_g$ -- that some internal degrees of freedom fall out of equilibrium with the heat bath. The system is then aged at $T_{\ell}$, for times insufficient to reach thermal equilibrium, and finally is heated abruptly and held at a temperature $T_f$ such that $T_{\ell}\! < \!T_f \!<\! T_h$.  The crucial observation is that, in this last stage of the Kovacs protocol, the volume  does not increase monotonically as a function of time, but goes through a maximum before decreasing slowly to its equilibrium value at $T_f$. The fact that the system can exist in two different states, on the upward and downward sides of the Kovacs volume peak, at the same temperature, pressure and volume, indicates that these are not states of thermal equilibrium. It is important to understand how to characterize them.

The Kovacs effect originally was observed in polyvinyl acetate \cite{KOVACS-63}, but since then it
has been observed in many other glassy polymers, see for example \cite{02BS} for measurements in polystyrene. Qualitatively similar memory effects have been observed in many other glassy systems such as colloidal glasses \cite{00BCL,03ONKDHM}, ferroelectrics \cite{01BDL-AL,01KB}, gelatin gels \cite{03PN}, granular materials \cite{00JTMJ}, superparamagnets and superspin glasses \cite{05SJT}. 
It also has been the subject of various recent theoretical investigations \cite{NIEUW-00,03BBDG, 03Buhot,04CLL,04AS,06ALN,LEUZZI-09}. 

In this paper, we look at the Kovacs effect from the point of view of our recent attempts to develop a first-principles, statistical formulation of nonequilibrium thermodynamics \cite{BLI-09,BLII-09,BLIII-09}. Generally speaking, our goal is to reinterpret the analysis of Kovacs {\it et al.} \cite{KOVACS-79} in terms of specific molecular processes.  Our work differs from that of Nieuwenhuizen, Leuzzi, and coworkers \cite{NIEUW-00,06ALN,LEUZZI-09}, for example, in that we start from a fundamental, statistical statement of the second law of thermodynamics and use it to derive equations of motion for relevant internal state variables as well as for an effective temperature.   We differ also from Bertin et al \cite{03BBDG}, who have solved specific models and have shown how phenomena analogous to the Kovacs effect emerge in interesting ways.

As in \cite{BLII-09}, our starting point is the assumption that a glass-forming material consists of two weakly interacting subsystems. The configurational (C) subsystem is specified by the set of mechanically stable molecular positions, that is, the inherent structures \cite{STILLINGER-WEBER-82,STILLINGER-88}. The kinetic-vibrational (K) subsytem is specified by all the other degrees of freedom -- the kinetic energies, the displacements of the molecules from their stable positions, and, in the case of more complex molecules such as those relevant to the Kovacs effect, the internal degrees of freedom of these molecules.  The physical rationale for this separation is the distinction between the time scales for dynamic processes in the two subsystems.
In situations where the system is driven by external forces, these two subsystems can fall out of thermodynamic equilibrium with each other.  The fast, K-subsystem remains in equilibrium with the heat bath at temperature $T$; the slower C-subsystem, at least transiently, has an effective temperature $\chi$ (in energy units) that is different from $k_B\,T$.

We propose that any model of the Kovacs effect should include three essential ingredients. First, and most obviously, we need to specify the configurational (C-subsystem) degrees of freedom that fall out of thermal equilibrium as the whole system is quenched into the vicinity of its glass temperature.  These  degrees of freedom must describe structural features that change via slow molecular rearrangements and thus do not keep up with more rapid variations of the bath temperature.  Since we are interested in volume changes, the simplest choice of this internal variable is a population of vacancy-like defects; but other internal degrees of freedom that couple to the volume might serve our purposes equally well.

Second, we need to ensure that the equations of motion for this out-of-equilibrium but statistically significant defect population are consistent with the laws of thermodynamics.  We have argued in \cite{BLII-09} that the natural way to do this is to use the effective temperature of the configurational subsystem, in direct analogy with Gibbsian statistical mechanics, to determine the states of maximum probability through which the configurational degrees of freedom are moving. These configurational degrees of freedom have well defined energies $U_C$ and entropies $S_C$; thus they have an effective temperature $\chi \!=\! \partial U_C/\partial S_C$, and their equations of motion must be based on their effective thermodynamics.

Third, and least obviously, we need mechanisms by which variations of the ordinary temperature of the K-subsystem can produce changes in the effective temperature of the C-subsystem. This means that we must understand how, and under what circumstances, ordinary thermal expansion and contraction become irreversible phenomena that can increase the entropy of the system as a whole.  Our nonequilibrium thermodynamic formulation suggests that there are two distinct thermo-viscoelastic mechanisms that can be relevant here. The first is a Kelvin-Voigt-type mechanism in which the K-subsystem exhibits a bulk viscosity arising directly from fast molecular interactions. The second is a somewhat slower Maxwell-type mechanism involving volume-related internal degrees of freedom that equilibrate directly with the ordinary temperature $T$ rather than the effective temperature $\chi$.  (Our main reference for models of thermo-viscoelasticity is Maugin's book on {\it The Thermomechanics of Nonlinear Irreversible Behaviors}, \cite{MAUGIN-99}). 

The work Mossa and Sciortino \cite{MS-04} offers a unique opportunity to test the ideas described above. These authors performed molecular dynamics simulations of a Kovacs experiment using the Lewis and Wahnstr\"{o}m model \cite{LW-94} of ortho-terphenyl (OTP), in which the molecules are rigid isosceles triangles interacting via a Lennard-Jones potential.  Their crucial result is that, when $T$ is increased from $T_{\ell}$ to $T_f$, both the volume and the inherent structure energy increase and go through maxima. This result tells us that the effective temperature and the vacancy population also increase and go through maxima, and that the thermal expansion driven by the change in the bath temperature is partially irreversible. More precisely, during thermal expansion or contraction at nonzero pressure, the system exchanges mechanical energy with its surroundings. Some of that energy is dissipated, producing configurational entropy.

One of the most remarkable results of Mossa and Sciortino \cite{MS-04} is shown in their Fig. 4.  There they demonstrate that, after the volume and the inherent structure energy go through maxima, the system can be described by states of quasi-equilibrium fully characterized by the effective temperature.  During the earlier stages, however, this is clearly not the case.  There, something else is happening that challenges our understanding of nonequilibrium thermodynamics.  That ``something else'' is the  central theme of the present investigation. 

The scheme of this paper is as follows.  In Sec. \ref{model}, we describe our two-subsystem model and comment on its physical ingredients.  The thermodynamic equations of motion for this model are derived in Sec. \ref{kovacs1}, where we show how the two mechanisms of irreversible thermo-viscoelasticity emerge from our nonequilibrium statistical analysis.  Section \ref{kovacs2} is devoted to identifying appropriate dimensionless variables and making first estimates of the parameters that appear in the scaled equations.

In Sec. \ref{MSdata}, we compare the predictions of our theory with the numerical simulations of Mossa and Sciortino \cite{MS-04}, and confirm that each of our three ingredients of a Kovacs model is, indeed, essential for understanding their data. During the reheating stage, as $T$ rises quickly from $T_{\ell}$ to $T_f$, both the fast Kelvin-Voigt-type and the somewhat slower Maxwell-type mechanisms are needed as sources of configurational entropy. These sources increase the volume on short and intermediate timescales and drive an increase in the effective temperature $\chi$. The volume continues to rise as the vacancy population grows toward a quasi-equilibrium value determined by the increased $\chi$. After this quasi-equilibrium is established, at about the time that the volume reaches its Kovacs peak, the vacancies remain in equilibrium with  $\chi$ as the system slowly ages toward a final state with $\chi \!=\! k_B\,T_f$.  In short, we recover the results of Mossa and Sciortino and fully agree with their interpretation of them.

\section{Ingredients of a Two-Subsystem Model}
\label{model}

Denote the total, extensive, internal energy of the two-subsystem model by
\begin{equation}
\label{Utot}
U_{tot} = U_C(S_C,V_{el},N_v)+ U_K(S_K,V_{el},N_a).
\end{equation}
The configurational (C-subsystem) energy, $U_C$, is a function of the configurational entropy $S_C$, an elastic volume $V_{el}$ that is common to both subsystems, and an extensive number of vacancy-like defects $N_v$ whose energies and excess volumes are $e_v$ and $v_v$ respectively.  Similarly, the kinetic-vibrational (K-subsystem) energy, $U_K$, is a function of the kinetic-vibrational entropy $S_K$, $V_{el}$, and the number  $N_a$ of what we call ``misalignment'' defects with energies $e_a$ and excess volumes $v_a$.

Assume that a fixed number of molecules, say $N_0$, occupies the elastic volume $V_{el}$, which does not include the excess volume of the defects. Then the total volume of the system is
\begin{equation}
\label{Vtot}
V_{tot} = V_{el} + N_v\,v_v + N_a\,v_a.
\end{equation}

Our picture of the vacancy-like defects in the C-subsystem is a slight oversimplification but seems conceptually simple.  In contrast, the misalignment defects require more discussion. The triangular geometry of an ortho-terphenyl molecule means that its volume and its energy depend on its orientation with respect to its neighbors.  Thus, even in the absence of the vacancy-like  defects that may characterize the slowly fluctuating configurational subsystem of OTP,  there are local misalignment defects that couple to the volume and can participate in energetically irreversible processes.  If the formation energies of these defects are not too large, and if the energy barriers that resist their transitions from one orientation to another are small enough, then these defects equilibrate quickly with the bath temperature and can legitimately be included in the kinetic-vibrational K-subsystem. For simplicity, we have assumed in Eqs. (\ref{Utot}) and (\ref{Vtot}) that there is only one kind of misalignment defect.

The assumption that the misalignment defects belong in the fast K-subsystem is not trivial.
Our model is similar to one studied by Ilg and Barrat \cite{I-B-07}, who show that the equilibration rate for a similar class of dynamical inclusions in a driven glass former depends sensitively on the strengths of the thermal noise sources that activate their transitions across internal barriers. We will show, however, that the high-temperature assumption works well for present purposes, and that these internal, orientational degrees of freedom produce a model of bulk thermo-viscoelasticity that is consistent with the MS data.

The temperature of the K-subsystem is
\begin{equation}
\label{thetadef}
\theta = k_B\,T =  \left({\partial U_K\over \partial S_K}\right)_{V_{el},N_a}.
\end{equation}
We assume that the K-subsystem, in addition to having its own internal dynamics, plays the role of a  thermal reservoir, so that $\theta$ is the temperature that is being controlled as a function of time during a Kovacs experiment. The effective temperature of the C-subsystem (in energy units) is
\begin{equation}
\label{chidef}
\chi = \left({\partial U_C\over \partial S_C}\right)_{V_{el},N_v}.
\end{equation}

Define the partial-pressure functions:
\begin{equation}
\label{pC}
p_C(\chi,V_{el}) = -\,\left({\partial F_C\over \partial V_{el}}\right)_{\chi,N_v};
\end{equation}
and
\begin{equation}
\label{pK}
p_K(\theta,V_{el})= -\,\left({\partial F_K\over \partial V_{el}}\right)_{\theta,N_a};
\end{equation}
where the free energies $F$ are the Legendre transforms of the $U$'s. Note, however, that we do not immediately identify $p_C \!+\! p_K$ as the total applied pressure. We do, however, assume strictly linear elasticity by writing
\begin{equation}
\label{FCelastic}
{F_C(\chi,V_{el})\over V_0} = {\lambda_C\over 2\,V_0^2}\,\Bigl[V_{el} - V_C(\chi)\Bigr]^2 + f_C(\chi),
\end{equation}
where $V_0$ is a reference volume, $\lambda_C$ is a compression modulus, $V_C(\chi)$ is the relaxed  C-subsystem volume, and $f_C(\chi)$ is a volume-independent free-energy density. Similarly,
\begin{equation}
\label{FKelastic}
{F_K(\theta,V_{el})\over V_0} = {\lambda_K\over 2\,V_0^2}\,\Bigl[V_{el} - V_K(\theta)\Bigr]^2 + f_K(\theta).
\end{equation}
The temperature dependent reference volumes  $V_C(\chi)$ and $V_K(\theta)$ are different from each other. The elastic energy of the C-subsystem has its minimum at a relatively small volume $V_C(\chi)$, because that system is cohesive at zero pressure. In contrast, the energy of the K-subsystem  decreases as the volume increases, because the kinetic energy is fixed and the vibrational modes become softer as the spacing between the molecules increases. Thus, strictly speaking, $V_K(\theta)$ must actually be defined at a positive reference pressure; but, since we are considering only linear elasticity, there is no need to be specific about this definition. In any case, the partial pressures defined in Eqs. (\ref{pC}) and (\ref{pK}) are different from each other and most likely have opposite signs, i.e. $p_C \!<\! 0\! <\! p_K$. This is indeed the case in the simulations of MS \cite{MS-04}.

\section{Thermodynamic Equations of Motion}
\label{kovacs1}

\subsection{First and Second Laws}

The first law of thermodynamics is
\begin{equation}
\label{firstlaw1}
-\,p\,\dot V_{tot} = \dot U_{tot},
\end{equation}
where $p$ is the applied pressure. With Eqs. (\ref{Vtot}), (\ref{thetadef}), (\ref{chidef}), (\ref{pC}) and (\ref{pK}), Eq.(\ref{firstlaw1}) becomes
\begin{eqnarray}
\label{firstlaw2}
\nonumber
\chi\,\dot S_C &+& \left[p\,v_v+\left({\partial U_C\over \partial N_v}\right)_{S_C,V_{el}}\right]\dot N_v +\, \theta\,\dot S_K \nonumber\\&+& \left[p\,v_a+\left({\partial U_K\over \partial N_a}\right)_{S_K,V_{el}}\right]\dot N_a \nonumber\\&+& \Bigl[p - p_C(\chi,V_{el}) - p_K(\theta,V_{el})\Bigr]\dot V_{el} = 0.
\end{eqnarray}
The second law is
\begin{equation}
\label{secondlaw1}
\dot S_{tot} = \dot S_C + \dot S_K \ge 0.
\end{equation}
Using Eq.(\ref{firstlaw2}) to eliminate $\dot S_C$, we write Eq.(\ref{secondlaw1}) in the form:
\begin{eqnarray}
\label{secondlaw2}
&-&\,\left[ p\,v_a + \left({\partial U_K\over\partial N_a}\right)_{S_K,V_{el}}\right]\,\dot N_a \\
&-&\,\left[ p\,v_0 + \left({\partial U_C\over\partial N_v}\right)_{S_C,V_{el}}\right]\,\dot N_v \nonumber\\ 
&-&\, \Bigl[p - p_C(\chi,V_{el}) - p_K(\theta,V_{el})\Bigr]\dot V_{el}-\, (\theta - \chi)\,\dot S_K \ge 0 \nonumber .
\end{eqnarray}
Following the procedure described in \cite{BLI-09,BLII-09}, we recognize that Eq.(\ref{secondlaw2}) consists of four separate inequalities associated with the independently variable quantities $\dot N_a$, $\dot N_v$, $\dot V_{el}$, and $\dot S_K$, and therefore we must satisfy four separate inequalities. In the next paragraphs, we look at these in reverse order of their appearance here.

\subsection{Aging Rate}

The inequality proportional to $\dot S_K$ is satisfied by writing
\begin{equation}
\label{C-heatflow}
\theta\,\dot S_K = -\,A(\chi,\theta)\,\left(1 - {\chi\over \theta}\right),
\end{equation}
where $A(\chi,\theta)$ is a non-negative thermal transport coefficient.  $\theta\,\dot S_K$  is the rate at which heat is flowing from the C-subsystem to the K-subsystem.  Since we assume that the coupling between the C and K-subsystems is weak, we expect $A(\chi,\theta)$ to be small.  This is the term that controls the rate at which the system ages in the absence of external driving.

\subsection{Kelvin-Voigt-Type Thermo-Viscoelasticity}

Next, consider the part of the inequality proportional to $\dot V_{el}$. Use the definitions of the partial pressures in Eqs. (\ref{pC}) and (\ref{pK}), plus the elastic free energies in Eqs. (\ref{FCelastic}) and (\ref{FKelastic}), to write
\begin{equation}
p - p_C(\chi,V_{el}) - p_K(\theta,V_{el}) = {\bar\lambda\over V_0}\,\bigl[V_{el}-V_{el}^{eq}(\chi,\theta,p)\Bigr],
\end{equation}
where $\bar\lambda = \lambda_C + \lambda_K$, and
\begin{equation}
\label{Veqdef}
V_{el}^{eq}(\chi,\theta,p) = \frac{1}{\bar\lambda}\,\Bigl[\lambda_K\,V_K(\theta)+ \lambda_C\,V_C(\chi) - p\,V_0\Bigr].
\end{equation}
The inequality in Eq.(\ref{secondlaw2}) is satisfied by writing an equation of motion for $V_{el}$:
\begin{equation}
\label{Veldot}
\tau_0\,\dot V_{el} = -\,\gamma\,\Bigl[V_{el}-V_{el}^{eq}(\chi,\theta,p)\Bigr],
\end{equation}
where $\tau_0$ is a molecular time scale and $\tau_0/\gamma$ is a bulk viscosity.  According to Maugin \cite{MAUGIN-99}, this is a Kelvin-Voigt-type thermo-viscoelasticity. When rewritten in terms of the pressures, Eq.(\ref{Veldot}) says that the driving force $p$ is equal to an elastic term, $p_C \!+\! p_K$,  plus a viscous force proportional to $\dot V_{el}$.

To see in more detail what is happening here, interpret the first-law in Eq.(\ref{firstlaw2}) as an equation for $\chi\,\dot S_C$, and note that the contribution to the configurational heating rate from the term proportional to $\dot V_{el}$ is
\begin{equation}
\label{Sdotel}
-\!\Bigl[p - p_C(\chi,V_{el}) - p_K(\theta,V_{el})\Bigr]\dot V_{el}\!=\!{\gamma\,\bar\lambda\over V_0}\,\bigl[V_{el}-V_{el}^{eq}(\chi,\theta,p)\Bigr]^2.
\end{equation}
Ordinarily, this term is negligible.  In the absence of a slow, internal, dissipative mechanism, the viscosity $\tau_0/\gamma$ is microscopically small. Suppose that some quantity on the right-hand side of Eq.(\ref{Veldot}), perhaps $p$ or $\theta$, is varied at an experimentally feasible rate, say $\nu/\tau_0 \!\ll\! \gamma/\tau_0$. By dimensional analysis of Eq.(\ref{Veldot}), we find that $\nu/\gamma \!\sim \!(V_{el}\!-\!V_{el}^{eq})/V_0$. If we then integrate the right-hand side of Eq.(\ref{Sdotel}) over a time of the order of $\nu^{-1}$, we find that the change in the total heat energy is of the order of $\bar{\lambda}\,V_0\,\nu/\gamma$, which vanishes when $\nu/\gamma \!\to\! 0$.  In this limit, the system becomes thermodynamically reversible. The solution of Eq.(\ref{Veldot}) is accurately $p \!=\! p_K \!+\! p_C$, which is the usual thermodynamic identity. In other words, we have recovered a special example of the general rule that equilibrium thermodynamics is valid when systems are driven quasistatically.

But the Kovacs effect is an exception to this rule. The original Kovacs observations were made with a glassy polymer, where the internal timescales $\tau_0/\gamma$ are long, so that it is possible to change temperatures and pressures relatively rapidly. It is also quite easy to do this in molecular dynamics simulations, which is what happens in the MS computations. We will see in Sec. \ref{MSdata} that the Kelvin-Voigt-type dissipation is an important driving force for the Kovacs effect.

\subsection{Maxwell-Type Thermo-Viscoelasticity}

We turn finally to the terms proportional to $\dot N_a$ and $\dot N_v$ in Eq.(\ref{secondlaw2}). These terms produce a Maxwell-type thermo-viscoelasticity, according Maugin \cite{MAUGIN-99}. To see this, look at the time derivative of the expression for the total volume in Eq.(\ref{Vtot}). If there is no Kelvin-Voigt-type viscous pressure proportional to $\dot V_{el}$, then the total deformation rate $\dot V_{tot}$ is the sum of an elastic term $\dot V_{el} = -\,V_0\,\dot p/\bar\lambda$, and a viscoelastic term $v_a\,\dot N_a+ v_v\,\dot N_v$. Our problem reduces to finding an expression for $v_a\,\dot N_a + v_v\,\dot N_v$ that will serve as a constitutive relation for the viscoelastic (viscoplastic) part of the deformation rate. To solve this problem, we follow steps outlined in \cite{BLI-09}.

Start with the K-subsystem. Assume that the entropy and energy of this system consist of separate, additive contributions, first from the defects and, second, from all the other degrees of freedom:
\begin{equation}
S_K(U_K,V_{el},N_a) = S_0(N_a) + S_1(U_1,V_{el});
\end{equation}
\begin{equation}
U_K(S_K,V_{el},N_a) = N_a\,e_a + U_1(S_1,V_{el});
\end{equation}
where, for simple defects without internal structure of their own,
\begin{equation}
\label{S0}
S_0(N) = N_0\,\ln N_0 -\,N\,\ln N -(N_0 -N)\,\ln(N_0 -\,N),
\end{equation}
and $N_0$, as defined earlier, is the total number of molecules. Then,
\begin{equation}
U_K(S_K,V_{el},N_a) = N_a\,e_a + U_1\Bigl[S_K - S_0(N_a),V_{el}\Bigr],
\end{equation}
so that
\begin{equation}
\left({\partial U_K\over \partial N_a}\right)_{S_K,V_{el}} = e_a + \theta\,{d S_0\over d N_a}.
\end{equation}

The inequality associated with the $\dot N_a$ term in Eq.(\ref{secondlaw2}) has the form of a Clausius-Duhem relation \cite{MAUGIN-99}, enforcing non-negative entropy production:
\begin{equation}
\label{CD1}
-\,\left[p\,v_a +  \left({\partial U_K\over \partial N_a}\right)_{S_K,V_{el}}\right] \dot N_a= -\,\left({\partial G_K\over \partial N_a}\right)_{\theta,p}\dot N_a \ge 0,
\end{equation}
where
\begin{equation}
G_K(\theta,p,N_a) = h_a\,N_a - \theta\,S_0(N_a);
\end{equation}
and $h_a = e_a + p\,v_a$. We satisfy Eq.(\ref{CD1}) by writing
\begin{equation}
\label{Nadot}
\tau_0\,\dot N_a = \Gamma_K\,\left[N_a^{eq}(\theta,p)-\,N_a\right],
\end{equation}
where  $\Gamma_K$ is a dimensionless rate factor and $\tau_0$ is the same molecular time scale that we introduced in Eq.(\ref{Veldot}). $N_a^{eq}(\theta,p)$ is the equilibrium value of $N_a$ determined by
\begin{equation}
\left({\partial G_K\over \partial N_a}\right)_{\theta,p}= 0 ~~{\rm at}~~ N_a = N_a^{eq}(\theta,p);
\end{equation}
which means that
\begin{equation}
N_a^{eq}(\theta,p)= {N_0\over e^{h_a/\theta}+1}.
\end{equation}

The same analysis pertains to the vacancy-like defects in the C-subsystem. The equation of motion for $N_v$ is the same as Eq.(\ref{Nadot}), but with $v_v$ and $e_v$ replacing $v_a$ and $e_a$, with $\chi$ instead of $\theta$, and with a new rate factor $\Gamma_C$:
\begin{equation}
\label{Nvdot}
\tau_0\,\dot N_v = \Gamma_C\,\left[N_v^{eq}(\chi,p)-\,N_v\right].
\end{equation}
We assume that the entropy associated with $N_v$ is the same function $S_0(N_v)$ that we introduced in Eq.(\ref{S0}); therefore
\begin{equation}
\label{Nveq}
N_v^{eq}(\chi,p)= {N_0\over e^{h_v/\chi}+1},
\end{equation}
where $h_v = e_v + p\,v_v$. We note that the idea that an internal variable can be transiently driven out of quasi-equilibrium with the effective temperature, as in Eq.(\ref{Nvdot}), was  introduced earlier in \cite{BLP07II}, where it was used to describe the internal dynamics of  deforming, simulated, amorphous silicon.

\subsection{Equation of Motion for the Effective Temperature $\chi$}

Putting these pieces together, we rewrite Eq.(\ref{firstlaw2}) as an expression for the heat flow into the C-subsystem:
\begin{eqnarray}
\label{firstlaw3}
\nonumber
\!\!\!&\chi&\!\!\,\dot S_C \!=\! -\left[h_v \!-\! \chi\,{\partial S_0(N_v)\over \partial N_v}\right]\,\dot N_v \!-\! \left[h_a - \theta\,{\partial S_0(N_a)\over \partial N_a}\right]\,\dot N_a\cr \\  &+& {\gamma\,\bar\lambda\over V_0}\,\bigl[V_{el}\!-\!V_{el}^{eq}(\chi,\theta,p)\Bigr]^2\!+\! A(\chi,\theta)\,\left(1 \!-\! {\chi\over\theta}\right).
\end{eqnarray}
The first three of these terms are non-negative rates of configurational heat production; the last term is the (ordinarily negative) rate at which heat flows from the K-subsytem into the C-subsystem. 

To convert this result into an equation of motion for $\chi$, write
\begin{eqnarray}
\nonumber
\chi\,\dot S_C &=& \chi\,{\partial S_C\over \partial \chi}\,\dot\chi + \chi\,{\partial S_C\over \partial N_v}\,\dot N_v + \chi\,{\partial S_C\over \partial V_{el}}\,\dot V_{el}\cr \\&=& C^{e\!f\!f}\,\dot\chi + \chi\,{\partial S_0\over \partial N_v}\,\dot N_v + \chi\,{\partial p_C\over \partial \chi}\,\dot V_{e}.
\end{eqnarray}
The second term exactly cancels the term proportional to $\partial S_0/\partial N_v$ on the right-hand side of Eq.(\ref{firstlaw3}), which becomes
\begin{eqnarray}
\label{chidot}
\nonumber
\!\!\!&C^{e\!f\!f}&\!\!\,\dot\chi = -\,h_v \,\dot N_v -\,\chi\,{\partial p_C\over \partial \chi}\,\dot V_{e}
 - \left[h_a - \theta\,{\partial S_0(N_a)\over \partial N_a}\right]\,\dot N_a\cr \\ &+&\! {\gamma\,\bar\lambda\over V_0}\,\bigl[V_{el}-V_{el}^{eq}(\chi,\theta,p)\Bigr]^2 \!+\! A(\chi,\theta)\,\left(1 \!-\! {\chi\over\theta}\right).
\end{eqnarray}

\section {Scaling and Approximations}
\label{kovacs2}

We have arrived at a complex set of equations with many variables and parameters. Before using these equations for data analysis, we rewrite them in terms of dimensionless quantities and, where possible, identify physically motivated estimates for some of the parameters.

To start, rescale the time so that $\tau_0 \!=\! 1$.  Then define the defect densities:
\begin{equation}
n_v = {N_v\over N_0};~~~~n_a = {N_a\over N_0};
\end{equation}
and the volume fractions:
\begin{equation}
\phi_{tot} = {V_{tot}\over V_0};~~~~\phi_{el}= {V_{el}\over V_0};~~~~\phi_{el}^{eq}= {V_{el}^{eq}\over V_0}.
\end{equation}
We also need to express the defect volumes in units of the volume per molecule:
\begin{equation}
\phi_v = {N_0\over V_0}\,v_v;~~~~\phi_a = {N_0\over V_0}\,v_a.
\end{equation}
Therefore
\begin{equation}
\label{phitot}
\phi_{tot} = \phi_{el} + \phi_a\,n_a + \phi_v\,n_v;
\end{equation}
and
\begin{equation}
\label{phieldot}
\dot \phi_{el} = -\,\gamma\,(\phi_{el}-\phi_{el}^{eq}).
\end{equation}

Measure $\chi$ in units of the vacancy enthalpy:
\begin{equation}
\tilde \chi = {\chi\over h_v};
\end{equation}
thus,
\begin{equation}
\label{nvdot}
\dot n_v = \Gamma_C\,\left({1\over e^{1/\tilde\chi}+1}- n_v\right)\approx \Gamma_C\,\left( e^{-1/\tilde\chi}- n_v\right),
\end{equation}
where the last approximation is valid in the low-density limit, $\tilde\chi\! \ll\! 1$.
For comparison with experimental data, it is convenient to express $\theta$ in units of absolute temperature $T$, so that
\begin{equation}
\dot n_a = \Gamma_K\,\left({1\over e^{T_a/T}+ 1}- n_a\right),
\end{equation}
where $k_B\,T_a \!\equiv\! h_a$.

For simplicity, assume that the relaxed volume of the C-subsystem, $V_C(\chi)$, is independent of $\chi$, and write
\begin{equation}
\phi_{el}^{eq}(T,p) \cong \phi_0 + \phi_1(T),
\end{equation}
where $\phi_0 \!=\! 1 \!-\! p/\bar\lambda$, and $\phi_1(T)$ describes the ordinary thermal expansion and contraction that drive the Kovacs experiment.  In general, $\phi_1(T)$ is a nonlinear function over the range of temperature jumps used by MS, and we will need to use their data to evaluate it.

The equation of motion for $\tilde\chi$, i.e. Eq.(\ref{chidot}), now reads
\begin{eqnarray}
\label{chidot2}
\nonumber
&& \tilde c^{e\!f\!f}\,\dot{\tilde\chi} = -\,\dot n_v - \beta\,\left[1 + {T\over T_a}\,\ln\,\left({n_a\over 1-n_a}\right)\right]\,\dot n_a \cr \\
&& +~ \gamma\,b\,\Bigl[\phi_{el}-\phi_{el}^{eq}(T,p)\Bigr]^2 + \Gamma_A(\tilde\chi,T)\,\left({T\over T_v} - \tilde\chi\right).~~~~~~~~
\end{eqnarray}
Here, $\tilde c^{e\!f\!f}\!=\!{C^{e\!f\!f}/ N_0}$, $k_B\,T_v \!=\! h_v,$ $\beta \!=\! h_a/h_v\!=\! T_a/ T_v$, $b \!=\! (\bar\lambda\,V_0/ h_v\,N_0)$ and $\Gamma_A(\tilde\chi,T) \!=\! A(\chi,\theta)/(N_0\,k_B\,T)$.\\

We can make rough estimates for some of these parameters using known properties of the OTP model simulated by MS. For example, taking parameters from their Lennard-Jones potential, we estimate the molecular vibration period to be about $1.5$ picoseconds.  Therefore, it is convenient to choose our unit of time to be $\tau_0 \!=\! 1$ ps.  Similarly, from the characteristic energy scale of this potential, we estimate that $h_v \sim 0.1$ eV, so that $T_v \!\sim\! 1300\,K$.  The misalignment defects must have substantially smaller formation energies.  If we guess that the difference is roughly a factor of ten, then $T_a \!\sim\! 130\,K$; so that these defects are far from being frozen out at the lowest temperatures used by MS, i.e. at $T_{\ell}\! = \!150\,K$.  This same estimate of $h_v$, combined with $V_0/N_0 \!\sim\! 0.4\, {\rm nm}^3$, and an estimate for the bulk modulus $\bar\lambda\!\sim\! 4$ GPa, tells us that $b\!\sim\! 50$, which means that the heating rate associated with the Kelvin-Voigt-type term in Eq.(\ref{chidot2}) may be substantial.

These estimates have interesting implications for our choices of the rate factors. Clearly, with $T_a \!\sim\! 130\,K$, the transition rate for the misalignment defects is not appreciably limited by an activation barrier; so $\Gamma_K$ must be only moderately slower than the molecular rate $\gamma$, which by definition cannot be significantly different from unity. Our first guess is that $\Gamma_K$ is in the range $10^{-2}$ to $10^{-1}$. On the other hand, $\Gamma_C$ should contain an effective thermal activation factor of the form $\exp\,(-\,\tilde\Delta_C/\tilde \chi)$, where $\tilde\Delta_C$ is the excess barrier, in units of $h_v$, that the system must surmount in either creating or annihilating a vacancy. If we assume that the system is fully equilibrated at $T \!=\! T_h \!=\! 400\,K$, then the initial value of $\tilde\chi$ is equal to $k_B\,T_h/h_v \!\sim\! 0.33$. Assuming that $\tilde\Delta_C$ is not too much smaller than unity, we conclude that $\Gamma_C$ may be smaller than $\Gamma_K$ by a factor of ten or so.

The more interesting rate factor is $\Gamma_A$, which, unlike $\Gamma_C$ in Eq.(\ref{nvdot}), is not the prefactor in a creation-rate formula that already contains an activation factor $\exp\,(-\,1/\tilde\chi)$. In other circumstances, such as a calculation of the $\alpha$ relaxation rate or the shear viscosity in a glass forming material that is moving so slowly that $\chi\!\approx\! k_B T$,  the analog of $\Gamma_A$ would be a super-Arrhenius function of $T$.  Here, although we are talking about the slow aging part of a Kovacs experiment, we are looking at the early transient stage where $\tilde\chi$ is still somewhat bigger than $T/T_v$ in Eq.(\ref{chidot2}). Accordingly, we assume that this rate at which the two weakly coupled subsystems equilibrate with each other is limited by a substantial energy barrier, say $\tilde\Delta_A \!\ge\! 1$; and we write
\begin{equation}
\label{GammaA}
\Gamma_A(\tilde\chi,T) \cong \Gamma_A^{(0)}(T)\,e^{-\,\tilde \Delta_A/\tilde\chi}.
\end{equation}
If the super-Arrhenius analogy is valid, $\Gamma_A^{(0)}(T)$ will be a rapidly varying function of $T$ that becomes vanishingly small below $T_g$. At constant $T$, while $\tilde\chi \!>\! T/T_v$, Eq.(\ref{GammaA}) implies that the aging rate slows exponentially as $\tilde\chi$ decreases.

\section{Comparisons with the data of Mossa and Sciortino}
\label{MSdata}

In their molecular dynamics experiments, MS are able to resolve dynamic variations in their model of ortho-terphenyl on time scales as small as tens of picoseconds.  In addition to observing volume changes, they can observe the energy, the pressure, and the shape factor (a measure of the width of the energy basins) of the inherent structures throughout the Kovacs protocol. Thus, they probe the Kovacs phenomena to a depth that seems impossible for laboratory experiments.

There are, however, compromises that must be made in such a procedure. MS simulate a system of only 343 OTP molecules.  Although they average their results over hundreds of initial configurations, it is hard to rule out effects of numerical noise, especially in the low-temperature aging calculations that must be dominated by very rare events.

Moreover, to control temperature and pressure, MS use a thermostat and a barostat with a time constant of $20$ ps; and they state that their systems are too far out of equilibrium for the data to be meaningful on shorter time scales following the initial quench or the reheating step. In spite of these uncertainties, we decided to try to model the complete Kovacs data reported by MS. We computed the values of the volume, the effective temperature, and the defect densities at the end of the aging stage, and used these values as the initial conditions for the reheating stage. Note that the interesting features of the Kovacs effect are very small; i.e. the fractional volume change associated with changes in $\chi$ near the Kovacs peak is only of the order of $10^{-3}$. Therefore, in some places, we have adjusted the values of our parameters to three or more significant figures in order to make quantitative comparisons with the MS data.

Our data fitting procedure for the results presented here started by choosing
\begin{equation}
\phi_{el}^{eq}(T= 400\,K)= \phi_0 = 1 - p/\bar\lambda = 0.9947,
\end{equation}
which was based on the estimate $p \!=\! 16\,{\rm MPa}$ and $\bar\lambda \!=\! 3\,{\rm GPa}$. We then estimated $T_v \!=\! 1300\,K$, $T_a \!=\! 130\,K$, $v_v \!=\! 0.07\,{\rm nm}^3$, and $v_a \!=\! 0.007\,{\rm nm}^3$. Assuming full equilibrium at $T_h \!=\! 400\,K$, $\tilde\chi \!=\! T_h/T_v$, we computed $n_v$ and $n_a$ at that temperature. MS report that their total volume per molecule at $400\,K$ is $0.378\,{\rm nm}^3$. These numbers uniquely determine $V_0/N_0 \!=\! 0.374\,{\rm nm}^3$.

From here on, we chose parameters in accord with our rough estimates at the end of Sec. \ref{kovacs2}, and refined these estimates to improve the agreement with the data. In addition to those cited in the last paragraph, the following numbers were used throughout the calculations: $b \!=\! 30$, $\tilde\Delta_A \!=\! 3.5$, $\Gamma_K \!=\! 10^{-\,1.75}$, and $\Gamma_C = 10^{-\,2.25}$. Our best-fit values for the elastic volume fractions at $T\!=\! 150\,K$ and $280\,K$ were $\phi_{el}^{eq}(T\!=\! 150\,K) \!=\! 0.9037$ and $\phi_{el}^{eq}(T\!=\! 280\,K)\!=\! 0.9323$.

\begin{figure}
\centering \epsfig{width=.52\textwidth,file=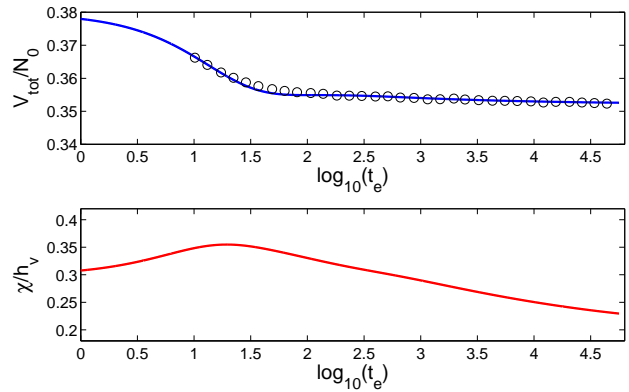} \caption{Aging at $T_{\ell}\!=\!280\,K$ after an instantaneous quench from $T_h\!=\!400\,K$. Upper panel: time evolution of the theoretical volume per molecule $V_0/N_0$ (solid line) compared to the simulation data (open circles) extracted from MS Fig. 1(c) \cite{MS-04}. The parameters used for the theoretical curve can be found in the text. Lower panel: the corresponding reduced effective temperature $\tilde\chi\!=\!\chi/h_v$.} \label{aging_280_a}
\end{figure}

\begin{figure}
\centering \epsfig{width=.52\textwidth,file=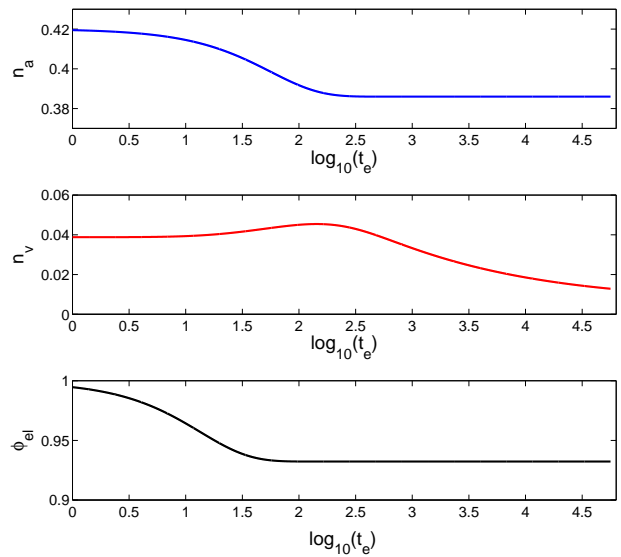} \caption{$n_a$, $n_v$ and $\phi_{el}$ corresponding to Fig. \ref{aging_280_a}.} \label{aging_280_b}
\end{figure}

In Figs. \ref{aging_280_a} and \ref{aging_280_b}, we show theory and data for the MS instantaneous quench from $T_h \!=\! 400\,K$ directly to $T_f \!=\! 280\,K$ and subsequent aging as functions of the time after quench $t_e$. (All times are stated in picoseconds.) The theoretical parameters were $\gamma \!=\! 10^{-\,1.13}$ and  $\Gamma_A^{(0)} \!=\! 10^{1.8}$. We need a Kelvin-Voigt-type viscosity with a small $\gamma$ because the elastic part of the volume initially relaxes on a time scale of the order of  $\gamma^{-\,1} \!\sim\! 15\,{\rm ps}$. The resulting dissipation drives a rapid increase in $\chi$; and then both $n_a$ and $n_v$ participate in the change of the total volume on time scales determined, respectively, by $\Gamma_K$ and $\Gamma_C$.

\begin{figure}
\centering \epsfig{width=.52\textwidth,file=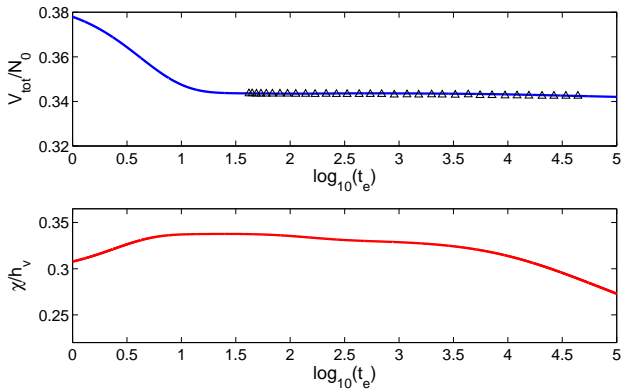} \caption{The same as Fig. \ref{aging_280_a}, but for $T_{\ell}\!=\!150\,K$, after a smooth quench from $T_h\!=\!400\,K$ according to Eq.(\ref{smooth_quench}). See text for the parameters used. The simulation data extracted from MS Fig. 1(c) \cite{MS-04}.} \label{aging_150_a}
\end{figure}

\begin{figure}
\centering \epsfig{width=.52\textwidth,file=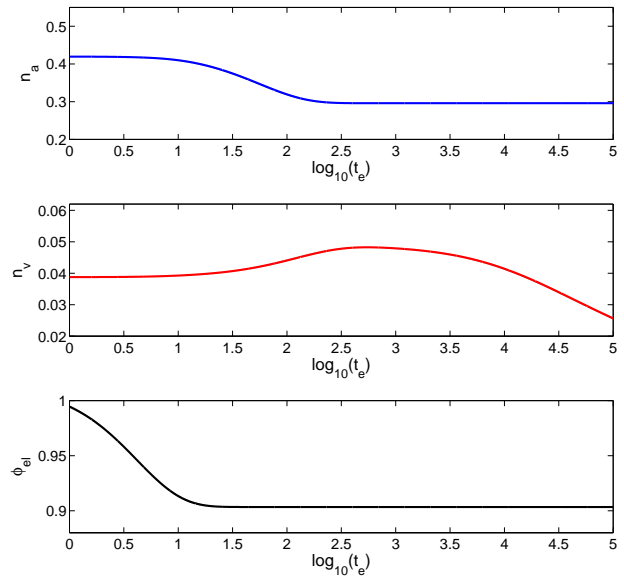} \caption{$n_a$, $n_v$ and $\phi_{el}$ corresponding to Fig. \ref{aging_150_a}.} \label{aging_150_b}
\end{figure}

Next consider the quench from $T_h \!=\! 400\,K$ to $T_{\ell} \!=\! 150\,K$ and subsequent aging at the latter temperature. In this case, MS have told us that they used a smooth, thermostatically controlled decrease in the temperature, and started to measure the volume at about $30$ ps after the quench was started \cite{private}. Accordingly, we have shifted our time scale by $30$ ps; and we have modeled the initial temperature dependence by writing
\begin{equation}
\label{smooth_quench}
T(t_e) = T_{\ell} +  (T_h-\,T_{\ell})\,\exp\,\left(-{t_e\over\tau_{th}}\right),
\end{equation}
with $\tau_{th} \!=\! 4\,{\rm ps}$. To use this equation for values of $T$ between $T_h$ and $T_{\ell}$, we have made a linear interpolation of $\phi_{el}^{eq}(T)$ between the values given above for $\phi_{el}^{eq}(T_h)$ and $\phi_{el}^{eq}(T_{\ell})$. We used $\gamma \!=\! 1$ and $\Gamma_A^{0} \!=\! 10^{-\,0.2}$. The results are shown in Figs. \ref{aging_150_a} and \ref{aging_150_b}, along with the MS data for the volume. With the more gradual quench and the larger value of $\gamma$, the Kelvin-Voigt-type effect is less pronounced but still present. The important feature here is the very slow aging at long times, associated with the smaller value of $\Gamma_A^{0}$ and the controlling effect of the effective-temperature activation barrier $\tilde\Delta_A$.

\begin{figure}
\centering \epsfig{width=.52\textwidth,file=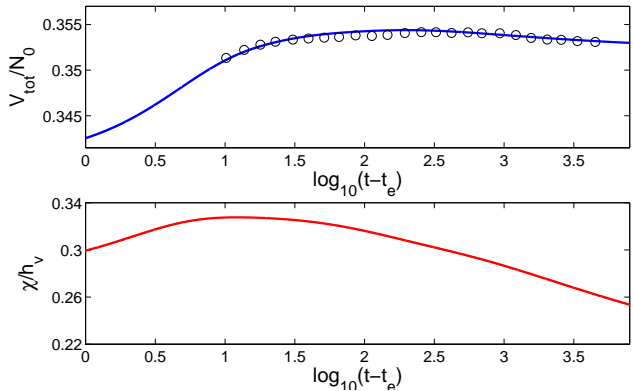} \caption{The same as Fig. \ref{aging_280_a}, but for reheating from $T_{\ell}\!=\!150\,K$ to $T_f\!=\!280\,K$ according to Eq.(\ref{smooth_reheating}), and after an aging time $\log_{10}(t_e)\!=\!4.4$ (equivalent to the nominal MS value of $25\,{\rm ns}$ on our shifted time scale). See text for the parameters used. The data in the upper panel are extracted from MS Fig. 2 \cite{MS-04}. The Kovacs peak appears to be small because the vertical scale includes the initial thermal expansion.  See Fig. \ref{reheating_280_c} for a closer look at the peak.} \label{reheating_280_a}
\end{figure}

\begin{figure}
\centering \epsfig{width=.52\textwidth,file=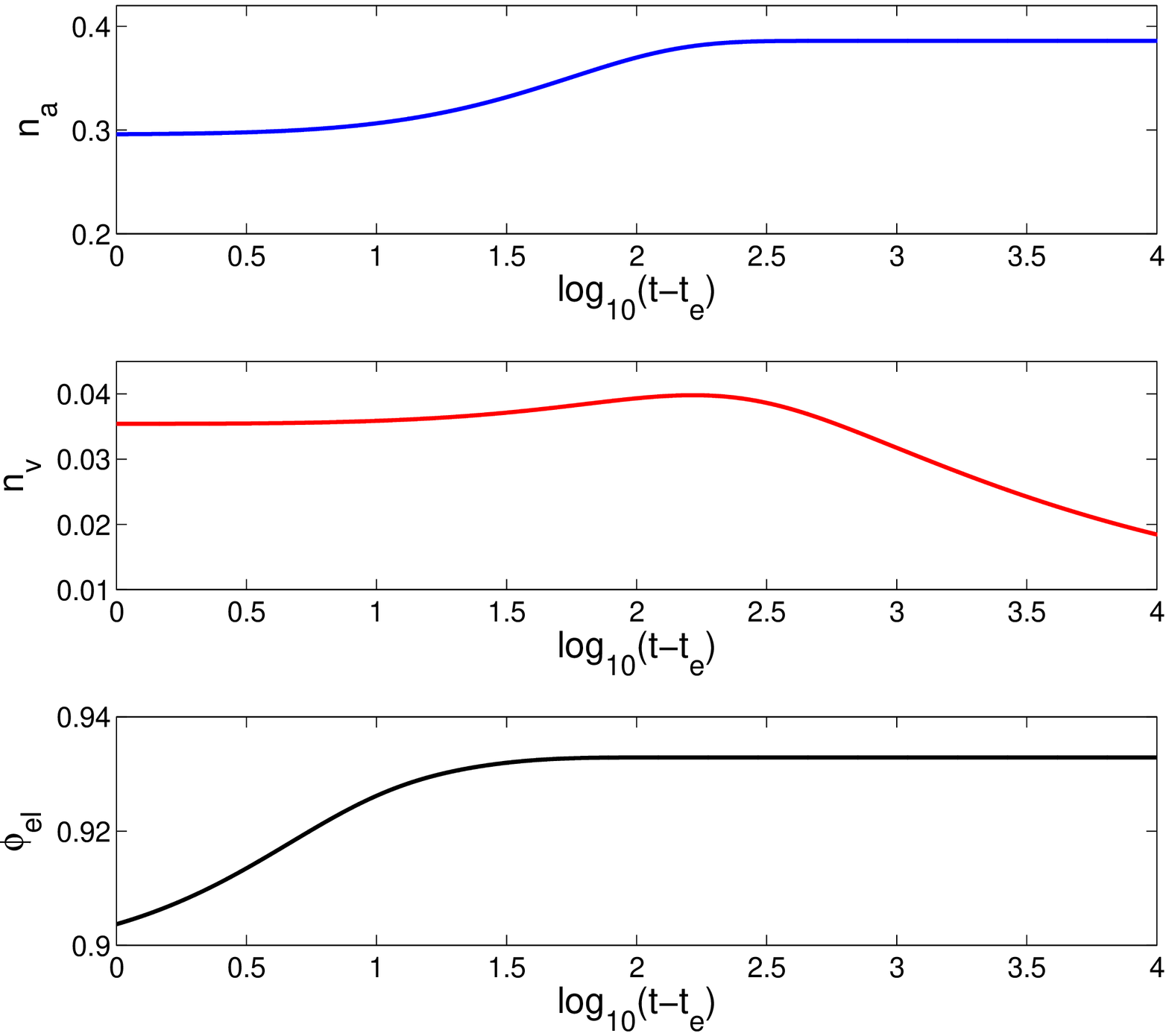} \caption{$n_a$, $n_v$ and $\phi_{el}$ corresponding to Fig. \ref{reheating_280_a}.} \label{reheating_280_b}
\end{figure}

The principal Kovacs effect occurs during and after reheating from $T_{\ell} \!=\! 150\,K$ to $T_f \!=\! 280\,K$, starting at the end of the aging period $t \!=\! t_e$. Because MS increased $T$ smoothly during reheating, we have used
\begin{equation}
\label{smooth_reheating}
T(t) = T_f - (T_f - T_{\ell})\,\exp\,\left(- {t - t_e\over \tau_{th}}\right),
\end{equation}
again with $\tau_{th} \!=\! 4\,{\rm ps}$, and with a linear interpolation between the lower and upper values of $\phi_{el}^{eq}(T)$. For this stage, we have used the same values of $\gamma$ and $\Gamma_A^{(0)}$ that we used for aging at $280\,K$. The results for a waiting time of $t_e \!=\! 25\,{\rm ns}$ (equivalent to the nominal MS value of $25\,{\rm ns}$ on our shifted time scale) are shown in Figs. \ref{reheating_280_a} and \ref{reheating_280_b}, along with the MS data for the volume. Again, we find that we need all three irreversible mechanisms to understand the observed behavior. The effective temperature increases quickly, due to the combination of both Kelvin-Voigt-type and Maxwell-type mechanisms. The density of K-subsystem defects, $n_a$, rises toward equilibrium with $T_f$ at a rate $\Gamma_K$; then the vacancy density $n_v$ rises toward equilibrium with $\chi$ at a rate $\Gamma_C$; and finally $n_v$, now at its quasi-equilibrium value as a function of $\chi$, decreases as $\chi$ decreases slowly toward $k_B\,T_f$.

\begin{figure}
\centering \epsfig{width=.52\textwidth,file=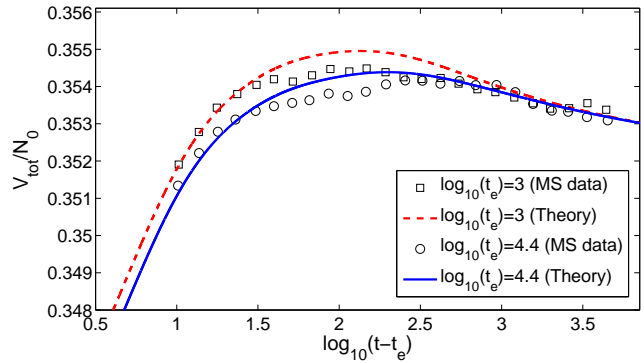} \caption{The same as the upper panel of Fig. \ref{reheating_280_a}, but after an aging time $\log_{10}(t_e)\!=\!3$ (equivalent to the nominal MS value of $1\,{\rm ns}$ on our shifted time scale). The upper panel of Fig. \ref{reheating_280_a} is copied here for comparison. See legend for more details.} \label{reheating_280_c}
\end{figure}

Figure \ref{reheating_280_c} shows the Kovacs peak from Fig. \ref{reheating_280_a} and the comparable peak for a shorter waiting time, $t_e \!=\! 1\,{\rm ns}$, along with the MS data for both cases. The agreement seems excellent in view of the fact that we computed the second curve only after having determined all of the parameters from the preceding calculations. 

\begin{figure}[here]
\centering \epsfig{width=.52\textwidth,file=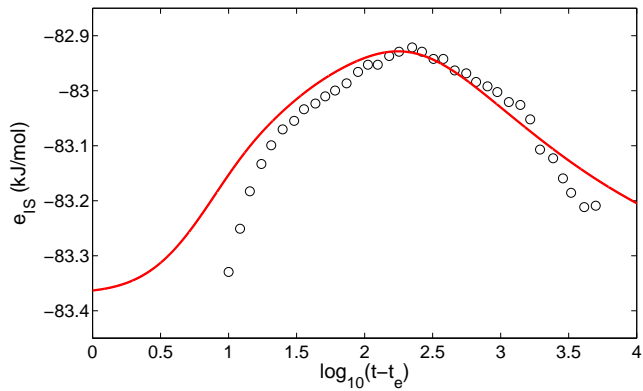} \caption{The theoretical inherent structure energy $e_{IS}$ of Eq.(\ref{IS_energy}) (solid line) compared to the simulation data (open circles) extracted from MS Fig. 3(b).} \label{e_IS}
\end{figure}

Finally, in Fig. \ref{e_IS}, we show the inherent-structure energy $e_{IS}$ for the reheating stage shown in Fig. \ref{reheating_280_a}. To fit the MS data, we have used
\begin{eqnarray}
\label{IS_energy}
e_{IS} &\cong& {\bar\lambda\,V_0\over 2\,N_0}\,\Bigl[\phi_{el}-\phi_{el}^{eq}(T_{\ell})\Bigr]^2 +n_a k_B\,T_a + n_v\,k_B\,T_v \nonumber\\
&+& \bar e_{IS} + \bar e_{IS}'\,k_B\,T_v\,\tilde\chi \ ,
\end{eqnarray}
with $\bar e_{IS} \!=\! -\,84.32$ kJ/mol and $\bar e_{IS}'\!=\! 0.08$. Note that our  inherent-structure energy $e_{IS}$ contains not only $U_C(S_C,V_{el},N_v)$, but also a contribution  from the misalignment defects $N_a$ that we earlier argued belong to $U_K$ for thermodynamic reasons. The conventional, static definition of the inherent-structure energy requires this interpretation of $e_{IS}$. We use the canonical variables $T$ and $\chi$ in Eq.(\ref{IS_energy}) instead of the entropies that we used in the micro-canonical formulation in Eq.(\ref{Utot}). Interestingly, the position of the peak in $e_{IS}(t)$ seems to be determined most strongly by the time dependence of the configurational vacancy term, i.e. $n_v\,k_B\,T_v$, as opposed to being determined predominantly  by other configurational degrees of freedom and thus more directly related to the time dependence of $\tilde\chi$.

\section{Concluding Remarks}

In their own concluding remarks, Mossa and Sciortino \cite{MS-04} summarize their results by saying that, instead of moving a system along a sequence of quasi-equilibrium configurations, their ``aging dynamics propagates the system through a sequence of configurations never explored in equilibrium, and it becomes impossible to associate the aging system to a corresponding liquid configuration.'' They go on to ask whether ``a thermodynamic description can be recovered [by] decomposing the aging system in a collection of substates, each of them associated with a different fictive $T$ ... or if the glass ... is trapped in some highly stressed configuration which can never be associated with a liquid state.''

We seem to be arriving at a related but different interpretation. By focusing on internal state variables -- in this case, the density of different kinds of defects in both the configurational and kinetic-vibrational subsystems -- in addition to the effective temperature, we naturally generate states in which the system as a whole departs from both ordinary thermal equilibrium with the bath temperature and from quasi-equilibrium with the effective temperature.  Our technique for making this calculation is the one we described in \cite{BLI-09,BLII-09}. We think that this technique goes at least part of the way toward answering the questions posed by Mossa and Sciortino, but we recognize that it encounters conceptual problems that eventually must be addressed.

The most obvious such problem, in our opinion, is the one that we found when deciding to include the misalignment defects among the fast degrees of freedom in the kinetic-vibrational subsystem. The results of Ilg and Barrat \cite{I-B-07} imply that some such internal variables may be neither completely fast nor completely slow but, rather, their dynamics might be activated by an intermediate temperature or noise strength.  This possibility might be loosely related to the MS conjecture about ``different fictive [temperatures];'' but the conjectures are intrinsically different from each other.  Neither we nor Ilg and Barrat are contemplating more than one effective temperature.  So far as we can tell, no complication of either kind is needed for understanding the Kovacs effect as observed by MS, nor do we seem to need it for shear flow in amorphous systems \cite{BLIII-09} or in polycrystals \cite{LBL-dislocations}.  Nevertheless, we appear to be encountering  some of the deepest and most important open questions in nonequilibrium physics.

\begin{acknowledgments}
We thank Stefano Mossa and Francesco Sciortino for their remarkably incisive work, for sharing their data with us, and for answering our questions about it.  We also thank Kenneth Kamrin for pointing out an error in our earlier thermodynamic analyses, which, had it not been corrected, would have given us a wrong result in Eq.(\ref{chidot}) in the present paper.  
\end{acknowledgments}

\end{document}